\documentclass[letterpaper, 10 pt, conference]{ieeeconf}  

\IEEEoverridecommandlockouts                              

\title{\LARGE \bf
A Comparative Study of Sensitivity Computations \\ in ESDIRK-Based Optimal Control Problems}

\author{Anders Hilmar Damm Andersen and John Bagterp Jørgensen
\thanks{A. H. D. Andersen and J. B. Jørgensen are with the Department of Applied
Mathematics and Computer Science, Technical University of Denmark, DK-2800 Kgs. Lyngby, Denmark.
Corresponding author: J. B. Jørgensen (E-mail: jbjo@dtu.dk).
        {\tt\small jbjo@dtu.dk}}%
}

\begin{document}

\maketitle
\thispagestyle{empty}
\pagestyle{empty}

\begin{abstract}
In this paper, we compare the impact of iterated and direct approaches to sensitivity computation in fixed-step explicit singly diagonally-implicit Runge–Kutta (ESDIRK) methods when applied to optimal control problems (OCPs). We use the principle of internal numerical differentiation (IND) strictly for the iterated approach, i.e., reusing the iteration matrix factorizations, the number of Newton-type iterations, and Newton iterates, to compute the sensitivities. The direct method computes the sensitivities without using the Newton schemes.
We compare the impact of the iterated and direct sensitivity computations in OCPs for the quadruple tank system. We benchmark the iterated and direct approaches with a base case. This base case is an OCP that applies an ESDIRK method that refactorizes the iteration matrix in every Newton iteration and uses a direct approach for sensitivity computations. In these OCPs, we vary the number of integration steps between control intervals and we evaluate the performance based on the number of SQP and QPs iterations, KKT violations, and the total number of function evaluations, Jacobian updates, and iteration matrix factorizations. The results indicate that the iterated approach outperforms the direct approach but yields similar performance to the base case.
\end{abstract}


\section{INTRODUCTION}

Efficient computation of integrator sensitivities is essential for gradient-based optimization algorithms applied to systems described by ordinary differential equations (ODEs). Multiple shooting-based optimal control problems (OCPs) belong to this category. They require the numerical solution of the system dynamics and the corresponding integrator sensitivities. One-step methods such as Runge-Kutta (RK) integration schemes have proven to be particularly effective when dealing with frequent discontinuities inherent in OCPs that implement zero-order hold parameterizations for the inputs \cite{Kristensen2004}.

For systems described by stiff ODEs or differential-algebraic equations, a multiple shooting-based OCP algorithm may require implicit Runge-Kutta (IRK) methods. Explicit singly diagonally-implicit Runge–Kutta (ESDIRK) integrators are such types of methods and we characterize them by a diagonal structure with identical coefficients on the diagonal and with the first stage being explicit. We can implement these methods such that they reuse the matrix factorizations for all Newton-type iterations within a single integrator step. We refer to this matrix as the \textit{iteration matrix}, and its reuse leads to a computationally efficient method.
ESDIRK-based multiple shooting OCPs have been applied to the continuous stirred tank reactor (CSTR) and the quadruple tank system (QTS) in
\cite{Capolei2012}.

There exist numerous approaches for sensitivity computations such as the \textit{Staggered direct method}, the \textit{Simultaneous corrector method}, and the \textit{Staggered corrector method}  \cite{Kristensen2004}. The Staggered direct method computes the sensitivity solutions directly from the iteration matrix after the system of nonlinear equations has converged. However, it requires frequent iteration matrix factorization to produce accurate integrator sensitivity information.
The Simultaneous corrector method solves a larger combined state and sensitivity equations system. This approach utilizes a special Jacobian structure and allows the factored iteration matrix to be reused for multiple steps while keeping a high accuracy of the sensitivities. Finally, the Staggered corrector method applies a separate Newton-type scheme to obtain sensitivities, after converging the system of nonlinear equations  \cite{Feehery_et_al_97}.

The principle of \textit{internal numerical differentiation} (IND) provides a methodology for developing sensitivities that are closely related to the integrator code \cite{Bock1981, Albersmeyer2010}. The principle of IND involves computing the sensitivities by directly differentiating the discretization scheme generated adaptively from the integrator. A strict implementation of IND for IRK methods results in reusing not only the step-sizes of the integrator but also the iteration matrix factorizations, number of Newton-type iterations, and the sequence of Newton iterates to compute sensitivities. Such sensitivity computations are applied to backward differentiation formula (BDF) methods and \cite{Albersmeyer2010} refers to it as the \textit{iterated} IND. This iterated IND method is similar to the Staggered corrector method.

In an alternative IND approach to sensitivity computations, we may assume that we solve the system of nonlinear equations exactly at each integration step. This allows for direct computation of the sensitivities based on the iteration matrix factorizations without using the Newton-type scheme.  This is referred to as the \textit{direct} IND approach in \cite{Albersmeyer2010} and is identical to the Staggered direct method. 

Different sensitivity computations for ESDIRK methods have already been shown, e.g., \cite{Kristensen2004} applies the direct approach for sensitivity computation for ESDIRK34 and suggests reusing the Jacobians for all the internal stages for one integrator iteration. The Staggered corrector approach is applied to ESDIRK12, ESDIRK23, and ESDIRK34 methods in \cite{Capolei2012}. The direct method, unlike the iterated method, avoids using the Newton-type scheme for sensitivity computation, making it a more computationally efficient option. However, this efficiency comes at the cost of obtaining only approximate sensitivities. A computational overview of the iterated and direct approaches is shown in \cite{Albersmeyer2010} and \cite{Quirynen2012} compares various computational aspects such as LU factorizations and Jacobian updates for iterated and direct IND methods. However, there has been no clear demonstration of the impact of these sensitivities on gradient-based optimization.\\

In this paper, we compare the computational performance of using either an iterated or a direct approach for sensitivity computations in ESDIRK-based multiple shooting OCPs. We do this using a sequential quadratic program (SQP) solver and compare the performances in terms of the number of SQP and QP iterations, function evaluations, Jacobian updates, and the number of iteration matrix factorizations. Our comparison is based on repeatedly solving OCPs for a model of the QTS with varying numbers of integration steps between control intervals. Additionally, we evaluate both approaches against a base case, where a direct approach to sensitivity computation is employed, but with the iteration matrix being refactorized in all Newton-type iterations in the ESDIRK integration scheme. The study shows that the iterated method is similar to the base case in terms of computational performance while avoiding some Jacobian updates and refactorizations in the integrator. We also show that the direct approach may only achieve convergence in the optimizer if we perform many integration steps for a small sampling time of the controller.

The rest of the paper is organized as follows. Section \ref{chap:ESDIRK} introduces the ESDIRK integration method for ODEs. Section \ref{chap:Sensitivity} shows how we compute sensitivities for the ESDIRK methods using the iterated and direct IND approaches. 
In Section \ref{chap:OCP} we present the optimal control problem. Section \ref{chap:Numerical_experiments} shows numerical examples using the ESDIRK-based multiple shooting OCP with either a direct or iterated IND approach and the base case applied to the QTS.
Finally, Section \ref{chap:Conclusions} presents conclusions.

\section{THE ESDIRK METHODS FOR ODES}
\label{chap:ESDIRK}
We consider the initial value problem (IVP)
\begin{subequations}
\label{eq:IVP}
\begin{alignat}{2}
       \dot{x}(t) &= f(t, x(t), u(t), d(t)), \quad  t \in [t_0, \, t_f]\\
       x(t_0) &= x_0,
\end{alignat}
\end{subequations}
where $x(t)\in\mathbb{R}^{n_x}$ is the state vector, $u(t)\in \mathbb{R}^{n_u}$ is the vector of inputs, and $d(t)\in \mathbb{R}^{n_d}$ is the vector of disturbances. The ESDIRK integration scheme for solving \eqref{eq:IVP} is
\begin{subequations}
    \label{eq:rk_scheme}
\begin{alignat}{2}
    T_i &= t_k+c_ih,\label{subeq:nodes}\\
    X_i &= x_k+h\sum_{j=1}^{i}a_{ij}f(T_j, X_j, u, d), \label{subeq:stages}\\
    x_{k+1} &= x_k + h \sum_{i=1}^{s}b_{i}f(T_i, X_i, u, d), \label{subeq:advancing_method}\\
    \hat{x}_{k+1} &= x_k+ h\sum_{i=1}^{s}\hat{b}_if(T_j, X_j, u, d), \label{subeq:bhat}
\end{alignat}
\end{subequations}
where $T_i$  and $X_i$ are the internal nodes and stages at iteration $k$ for $i = 1, \dots, s$, with $s$ being the number of stages, and $x_k$ and $x_{k+1}$ are the steps computed at $t_k$ and $t_{k+1} = t_k + h$, respectively. We may compute the integration step-size $h$ based on the local error estimate, $e_{k+1} = x_{k+1}-\hat{x}_{k+1}$, using the embedded method in \eqref{subeq:bhat}. However, for simplicity, we only consider a fixed integration step-size selection in this paper. Here, $h$ is defined as $h = \frac{t_f-t_0}{N}$, where $N$ is the number of integration steps. 

Table \ref{tab:butcher_tab_ESDIRK} presents the Butcher tableaus for ESDIRK12, ESDIRK23, and ESDIRK34. We apply the numerical values for these tableaus from \cite{Jørgensen_2018}.  For the remaining part of the paper, we apply the simplified notation $f(X_i, u) = f(T_i, X_i, u(t), d(t))$. \\
\begin{table}[tb]
\addtolength{\tabcolsep}{-3pt}    
\caption{Butcher tableau's for some ESDIRK methods.}
\label{tab:butcher_tab_ESDIRK}
    \begin{minipage}[t]{.21\linewidth}
        \vspace{0pt} 
        \centering
        
        \begin{tabular}{ccc}
        \multicolumn{3}{c}{ESDIRK12}       \\ \\
        \multicolumn{1}{c|}{$0$} & $0$ &   \\
        \multicolumn{1}{c|}{$1$} & $b_1$ & $\gamma$ \\ \hline
        \multicolumn{1}{c|}{$x_{k+1}$} & $b_1$ & $\gamma$ \\ \hline
        \multicolumn{1}{c|}{$\hat{x}_{k+1}$} & $\hat{b}_1$ & $\hat{b}_2$ 
        \end{tabular}
    \end{minipage}
    \begin{minipage}[t]{.33\linewidth}
        \vspace{0pt} 
        \centering
        
        \begin{tabular}{cccc}
        \multicolumn{4}{c}{ESDIRK23}       \\ \\
        \multicolumn{1}{c|}{$0$} & $0$ &    \\
        \multicolumn{1}{c|}{$c_2$} & $a_{21}$ & $\gamma$ &  \\ 
        \multicolumn{1}{c|}{$1$} & $b_1$ & $b_1$ & $\gamma$ \\ \hline
        \multicolumn{1}{c|}{$x_{k+1}$} & $b_1$ & $b_1$ & $\gamma$ \\ \hline
        \multicolumn{1}{c|}{$\hat{x}_{k+1}$} & $\hat{b}_1$ & $\hat{b}_2$ & $\hat{b}_3$
        \end{tabular}
    \end{minipage}
    \begin{minipage}[t]{.35\linewidth}
        
        \vspace{0pt} 
        \centering
        
        \begin{tabular}{ccccc}
        \multicolumn{5}{c}{ESDIRK34}       \\ \\
        \multicolumn{1}{c|}{$0$} & $0$ &   \\
        \multicolumn{1}{c|}{$c_2$} & $a_{21}$ &  $\gamma$ &  \\ 
        \multicolumn{1}{c|}{$c_3$} & $a_{31}$ & $a_{32}$ & $\gamma$ &  \\ 
        \multicolumn{1}{c|}{$1$} & $b_1$ & $b_2$ & $b_3$ & $\gamma$\\ \hline
        \multicolumn{1}{c|}{$x_{k+1}$} & $b_1$ & $b_2$ & $b_3$ & $\gamma$ \\ \hline
        \multicolumn{1}{c|}{$\hat{x}_{k+1}$} & $\hat{b}_1$ & $\hat{b}_2$ & $\hat{b}_3$ & $\hat{b}_4$ 
        \end{tabular}
    \end{minipage}
    \addtolength{\tabcolsep}{3pt}
\end{table}
For each integration step, we solve the systems of nonlinear equations
\begin{equation}
\label{eq:ESDIRK_residual}
    R_i(X_i) = X_i-h\gamma f(X_i, u)-\psi_i = 0, 
\end{equation}
for $i = 2, \dots, s$, with 
\begin{equation}
\psi_i = \psi_i(\{X_j\}_{j=1}^{i-1}, x_k, u) = x_k+h\sum_{j=1}^{i-1}a_{ij}f(X_j, u).
\end{equation}
We do this by applying the inexact Newton method 
\begin{subequations}
    \label{eq:newton_method}
\begin{alignat}{2}
    M_k\Delta X_i^{[l]} &= -R_i(X_i^{[l]}),\\
    X_i^{[l+1]} &= X_i^{[l]} +\Delta X_i^{[l]},
\end{alignat}
\end{subequations}
for $l = 0, \dots, w_{i,k}-1$, with $w_{i,k}$ being the number of iterations required for stage $i$ at iteration $k$ to satisfy a chosen convergence criteria. $M_k$ is the iteration matrix at iteration $k$ defined as
\begin{equation}
\label{eq:iteration_matrix}
    M_k = I-h\gamma \frac{\partial f(x_k, u) }{\partial x_k}\approx \frac{\partial R_i(X_i^{[l]})}{\partial X_i^{[l]}},
\end{equation}
where $I$ is the identity matrix. We solve \eqref{eq:newton_method} using the LU factorizations of $M_k$ and we update these in every integration step. We choose the convergence criteria for the inexact Newton method as
\begin{equation}
\label{eq:stop_crit_newton}
    \left \| R_i(X_i^{[l]}) \right \| = \max_{j\in 1, \dots ,n_x}\frac{|R_i(X_i^{[l]})_j|}{\max(\text{abs}, \,\text{rel}(X_i^{[l]})_j)} < \tau,
\end{equation}
with abs and rel being absolute and relative tolerances, respectively, and $\tau = 0.1$, for $i=2,\dots, s$ \cite{Capolei2012}.
As the ESDIRK methods are stiffly accurate, i.e., $a_{sj} = b_j$, we avoid the computations in \eqref{subeq:advancing_method} and obtain the next step directly as
\begin{equation}
\label{eq:stiffly_accurate_advancing}
    x_{k+1} = X_s^{[w_{s,k}-1]}.
\end{equation}

\subsection{Computation of initial guesses for Newton iterations}

We apply stage value predictors (SVPs) to generate initial guesses of the Newton iteration schemes. The SVPs use information from the previously converged step, $x_{k-1}$, and previous converged stages, $\hat{X_j} = \hat{X_j}^{[w_{j,k-1}-1]}$ for $j=2, \dots s$, to construct guesses on the form
\begin{equation}
\label{eq:SVP}
    X_i^{[0]} = \alpha_i(r)x_{k-1} + \sum_{j=2}^{s}\beta(r)_{ij}\hat{X_j}, \;\text{for} \; i=2, \dots, s.
\end{equation}
$\alpha(r) \in \mathbb{R}^{s-1}$ and $\beta(r) \in \mathbb{R}^{s-1\times s-1}$ are the predictor coefficients computed using step-size ratio $r = h_{k}/h_{k-1}$. For fixed integration step-size $r=1$. We construct these predictors using the order conditions in \cite{Higeras_Roldan_2005}. The SVPs for ESDIRK12, ($\alpha^{12}$, $\beta^{12}$), and for ESDIRK23,  ($\alpha^{23}$, $\beta^{23}$) are
\begin{subequations}
\label{eq:SVP_ESDIRK_12_23}
\begin{alignat}{2}
\alpha^{12}(r) &= -r, \quad \beta^{12}(r) = 1+r, \\
    \alpha^{23}(r) &=\begin{bmatrix}
r - 2 \gamma r + 2\gamma r^2\\ 
\frac{r - 2\gamma r + r^2}{2\gamma}
\end{bmatrix},\\ \beta^{23}(r) &= \begin{bmatrix}
\frac{2 \gamma r^2 + r}{2\gamma  - 1} & \frac{-(4\gamma^2r^2 - 4\gamma^2r + 4\gamma r - 2\gamma  + 1)}{(2\gamma  - 1)} \\ 
\frac{r^2 + r}{(2\gamma (2\gamma  - 1))} & \frac{-(2r - 2\gamma  - 2\gamma r + r^2 + 1)}{(2\gamma  - 1)}
\end{bmatrix}.
\end{alignat}
\end{subequations}
We don't show the SVPs for ESDIRK34 due to the lack of space. For the first integration step ($k=0$), we use the trivial predictor, i.e., $X^{[0]}_i = x_k$. We note, that due to \eqref{eq:stiffly_accurate_advancing}, \eqref{eq:SVP} applies information of $x_k$ as $\hat{X_s}^{[w_{s,k-1}-1]}=x_k$.

\section{SENSITIVITY ANALYSIS}
\label{chap:Sensitivity}


\subsection{Iterated IND}
In the iterated IND, we differentiate the adaptively generated discretization scheme, including the operations in the Newton-type scheme. The adaptive components for the fixed-stepsize ESDIRK methods are the LU factorizations of the iteration matrices, $M_k$, the number of Newton-type iterations for all stages, $w_{i,k}$, and the sequence of Newton iterates, $X_i^{[l]}$ for $l = 0, \dots, w_{i,k}-1$ and for $i = 2, \dots, s$.  We express the implementation of the ESDIRK methods as a combination of the elementary operations
\begin{subequations}
    \label{eq:esdirk_elementary_operations}
\begin{alignat}{2}
    X_i^{[0]} &= \phi^{SVP}_i(x_{k-1}, \hat{X}), \label{subeq:initGuess}\\
    X_i^{[l+1]} &= \phi^{it}_i( X_i^{[l]}, \psi_i, u) \nonumber\\
    &= X_i^{[l]} - M_k^{-1}R_i(X_i^{[l]}, \psi_i, u),\label{subeq:NIter}\\
    x_{k+1} &= \phi^{s}(X_s^{[w_{s,k}-1]}) = X_s^{[w_{s,k}-1]}\label{subeq:stage2nextstate},
\end{alignat}
\end{subequations}
where \eqref{subeq:initGuess} represent the SVPs for stage $i$ in \eqref{eq:SVP} with $\hat{X}~=~\begin{bmatrix}
\hat{X}_2^{[w_{2,k-1}-1]}; & \dots; & x_k
\end{bmatrix}$, \eqref{subeq:NIter} represent the Newton-type iterations in \eqref{eq:newton_method}, and \eqref{subeq:stage2nextstate} represent the mapping of the final stage to the next step in \eqref{eq:stiffly_accurate_advancing}.  

\subsubsection{State sensitivity}
We construct the state sensitivities by computing the partial derivatives of \eqref{eq:esdirk_elementary_operations} with respect to the initial state, $x_0$, i.e.,
\begin{subequations}
    \label{eq:dev_esdirk_elementary_operations_state}
\begin{alignat}{2}
    \frac{\partial X_i^{[0]}}{\partial x_0} &= \frac{\partial \phi^{SVP}_i(x_{k-1}, \hat{X})}{\partial x_0}, \label{subeq:initGuess_dx0} \\
    \frac{\partial X_i^{[l+1]}}{\partial x_0} &= \frac{\partial \phi^{it}_i( X_i^{[l]}, \psi_i, u)}{\partial x_0}, \label{subeq:NIter_dx0}\\
   \frac{\partial x_{k+1}}{\partial x_0} &= \frac{\partial \phi^{s}(X_s^{[w_{s,k}-1]})}{\partial x_0} \label{subeq:stage2nextstate_dx0},
\end{alignat}
\end{subequations}
for $i = 2, \dots, s$, with \eqref{subeq:NIter_dx0} being
\begin{equation}
    \frac{\partial X_i^{[l+1]}}{\partial x_0} = \frac{\partial X_i^{[l]}}{\partial x_0}-M_k^{-1}\frac{\partial R_i(X_i^{[l]}, \psi_i, u)}{\partial x_0}.
\end{equation}
We compute the partial derivative of the residual function as
\begin{equation}
    \frac{\partial R_i(X_i^{[l]}, \psi_i, u)}{\partial x_0} = \frac{\partial R_i}{\partial X_i^{[l]}}\frac{\partial X_i^{[l]}}{\partial x_0} + \frac{\partial R_i}{\partial \psi_i}\frac{\partial \psi_i}{\partial x_0}
\end{equation}
with 
\begin{subequations}
\begin{alignat}{2}
    \frac{\partial R_i}{\partial X_i^{[l]}} &= \bigg(I-h\gamma\frac{\partial f(X_i^{[l]}, u)}{\partial X_i^{[l]}}\bigg), \quad
    \frac{\partial R_i}{\partial \psi_i} &= -I,\label{subeq:dRdXi_and_dRdpsi}\\
    \frac{\partial \psi_i}{\partial x_0} &= \frac{\partial x_k}{\partial x_0}+h\sum_{j=1}^{i-1}a_{ij}\frac{\partial f(X_j, u)}{\partial X_j}\frac{\partial X_j}{\partial x_0}, \label{subeq:psi_x0}
\end{alignat}
\end{subequations}
where $X_j = X_j^{[w_{j,k}-1]}$ for $j=1,\dots, i-1$ in \eqref{subeq:psi_x0} are the stages converged after $w_{j,k}$-number of Newton-type iterations at integrator iteration $k$. For \eqref{subeq:initGuess_dx0}, we compute the state sensitivities of the SVPs as 

\begin{equation}
    \frac{\partial X_i^{[0]}}{\partial x_0} =  
\begin{cases}
    \alpha_i\frac{\partial x_{k-1}}{\partial x_0} + \sum_{j=2}^{s} \beta_{ij}\frac{\partial \hat{X}_j}{\partial x_0},& \text{if } k\geq 1\\
    \frac{\partial x_k}{\partial x_0}, & \text{if } k =  0
\end{cases}
\end{equation}

for $i = 2, \dots, s$ and for \eqref{subeq:stage2nextstate_dx0} we write
\begin{equation}
\label{eq:Akp1}
    \frac{\partial x_{k+1}}{\partial x_0} = \frac{\partial X_s^{[w_{s,k}-1]}}{\partial x_0}.
\end{equation}

\subsubsection{Input sensitivity}
We construct sensitivities with respect to the input sensitivity as

\begin{subequations}
    \label{eq:dev_esdirk_elementary_operations_input}
\begin{alignat}{2}
    \frac{\partial X_i^{[0]}}{\partial u} &= \frac{\partial \phi^{SVP}_i(x_{k-1}, \hat{X})}{\partial u},\label{subeq:initGuess_du}\\
    \frac{\partial X_i^{[l+1]}}{\partial u} &= \frac{\partial \phi^{it}_i(X_i^{[l]}, \psi_i, u)}{\partial u}, \label{subeq:NIter_du}\\
   \frac{\partial x_{k+1}}{\partial u} &= \frac{\partial \phi^{s}(X_s^{[w_{s,k}-1]})}{\partial u}, \label{subeq:stage2nextstate_du}
\end{alignat}
\end{subequations}
for $i = 2, \dots, s$, We compute \eqref{subeq:NIter_du} as
\begin{equation}
    \frac{\partial X_i^{[l+1]}}{\partial u} = \frac{\partial X_i^{[l]}}{\partial u}-M_k^{-1}\frac{\partial R_i(X_i^{[l]}, \psi_i, u)}{\partial u},
\end{equation}
where the partial derivative of the residual function is
\begin{equation}
    \frac{\partial R_i(X_i^{[l]}, \psi_i, u)}{\partial u} = \frac{\partial R_i}{\partial X_i^{[l]}}\frac{\partial X_i^{[l]}}{\partial u} + \frac{\partial R_i}{\partial \psi_i}\frac{\partial \psi_i}{\partial u} + \frac{\partial R_i}{\partial u}.
\end{equation}
$\frac{\partial R_i}{\partial X_i^{[l]}}$ and $\frac{\partial R_i}{\partial \psi_i}$ are the same as in \eqref{subeq:dRdXi_and_dRdpsi} and we calculate $\frac{\partial \psi_i}{\partial u}$ and $\frac{\partial R_i}{\partial u}$ as

\begin{subequations}

\begin{alignat}{2}
    \frac{\partial \psi_i}{\partial u} &= \frac{\partial x_k}{\partial u}+h\sum_{j=1}^{i-1}\bigg(a_{ij}\frac{\partial f(X_j,u)}{\partial X_j}\frac{\partial X_j}{\partial u}+\frac{\partial f(X_j, u)}{\partial u}\bigg), \label{eq:psi_u}\\
   \frac{\partial R_i}{\partial u} &= -h\gamma\frac{\partial f(X_i^{[l]}, u)}{\partial u}.
\end{alignat}
\end{subequations}
The contribution to the sensitivities for the SVPs in \eqref{subeq:initGuess_du} is 
\begin{equation}
    \frac{\partial X_i^{[0]}}{\partial u} =  
\begin{cases}
    \alpha_i\frac{\partial x_{k-1}}{\partial u} + \sum_{j=2}^{s} \beta_{ij}\frac{\partial \hat{X}_j}{\partial u},& \text{if } k\geq 1\\
    \frac{\partial x_k}{\partial u}, & \text{if } k =  0
\end{cases},
\end{equation}
for $i = 2, \dots, s$. For \eqref{subeq:stage2nextstate_du} the input sensitivity at the next iteration is directly obtained as
\begin{equation}
\label{eq:Bkp1}
    \frac{\partial x_{k+1}}{\partial u} = \frac{\partial X_s^{[w_{s,k}-1]}}{\partial u}.
\end{equation}

We initialize the state and input sensitivities at integration iteration $k=0$ as
\begin{equation}
\label{eq:init_sensitivity}
    \frac{\partial x_{k}}{\partial x_0} = I, \quad \frac{\partial x_{k}}{\partial u} = 0.
\end{equation}
\subsubsection{Implementations details}
The ESDIRK integration schemes are implemented such that at least one Newton-type iteration is performed for each stage. This ensures that all elementary operations in \eqref{eq:esdirk_elementary_operations} are included when computing the sensitivities. 

\subsection{Direct IND}

In the direct IND approach, we assume we solve \eqref{eq:ESDIRK_residual} for $i = 2, \dots, s$, exactly. Consequently, this enables the direct computation of state and input sensitivities for \eqref{eq:rk_scheme} as
\begin{subequations}
\label{eq:direct_ind}
\begin{alignat}{2}
        \frac{\partial X_i}{\partial x_0} &= \frac{\partial x_k}{\partial x_0} + h\sum_{j=1}^{i}a_{ij}\frac{\partial f(X_j, u)}{\partial X_j}\frac{\partial X_j}{\partial x_0} \nonumber\\
        &= J_i^{-1}\frac{\partial\psi_i}{\partial x_0}, \label{subeq:base_base_sensitivity_x0}\\
        \frac{\partial X_i}{\partial u} &= \frac{\partial x_k}{\partial u} + h\sum_{j=1}^{i}a_{ij}\bigg(\frac{\partial f(X_j, u)}{\partial X_j}\frac{\partial X_j}{\partial u}+ \frac{\partial f(X_j, u)}{\partial u}\bigg) \nonumber\\
        &= J_i^{-1}\bigg(\frac{\partial\psi_i}{\partial u}+h\gamma\frac{\partial f(X_i,u)}{\partial u}\bigg),\label{subeq:base_base_sensitivity_u}
\end{alignat}
\end{subequations}
where $J_i=\frac{\partial R_i(X_i)}{\partial X_i}$, and $\frac{\partial\psi_i}{\partial x_0}$ and $\frac{\partial\psi_i}{\partial u}$ are described in \eqref{subeq:psi_x0} and \eqref{eq:psi_u}, respectively. We apply the approximation in \eqref{eq:iteration_matrix} to reuse the factorizations of $\frac{\partial R_i(X_i)}{\partial X_i}$ in \eqref{eq:direct_ind}
\begin{subequations}
\label{eq:direct_ind_approx}
\begin{alignat}{2}
        &\frac{\partial X_i}{\partial x_0} \approx M_k^{-1}\frac{\partial\psi_i}{\partial x_0},\label{subeq:approx_direct_x0}\\
        &\frac{\partial X_i}{\partial u}  \approx M_k^{-1}\bigg(\frac{\partial\psi_i}{\partial u}+h\gamma\frac{\partial f(X_i,u)}{\partial u}\bigg).\label{subeq:approx_direct_u}
\end{alignat}
\end{subequations}
Identically to the iterated approach, we obtained $\frac{\partial x_{k+1}}{\partial x_0}$ and $\frac{\partial x_{k+1}}{\partial u}$ as \eqref{eq:Akp1} and \eqref{eq:Bkp1}, respectively, and we apply the same initial values as in \eqref{eq:init_sensitivity}. We refer to \eqref{eq:direct_ind_approx} as the \textit{direct} IND approach for the ESDIRK methods \eqref{eq:rk_scheme}.

\subsection{Comments on iteration matrix refactorization strategies}

Some ODE/DAE solvers improve computational efficiency by reusing iteration matrix factorizations through multiple consecutive integration steps when solving \eqref{eq:IVP}. Refactorization strategies similar to the strategies in \cite{Gustafsson1997} have successfully been implemented for ESDIRK methods, as demonstrated in \cite{Voelcker2010}. However, integrators that apply such strategies in gradient-based optimization may experience convergence issues, especially those using the direct IND method. This is because factorizations are not only used for solving nonlinear equations but also for directly obtaining the sensitivities. In contrast, as suggested in \cite{Albersmeyer2010},  we can equip integrators using the iterated approach with such refactorization strategies. This makes the iterated IND approach an attractive choice for IRK methods, where these strategies significantly enhance efficiency. However, it is important to note that the iterated method still requires frequent Jacobian updates, which can potentially eliminate advantages gained from such a strategy.

\section{OPTIMAL CONTROL PROBLEM}
\label{chap:OCP}

We formulate an OCP as a weighted least-squares problem that minimizes the deviation between outputs and setpoints and penalizes the input rate of movement. We formulate this OCP from $t_0$ to $t_f = t_0+T$, where $T$ is the prediction horizon. We apply a zero-order holds parameterization of the inputs, i.e., $u(t) = u_n$ for $t_n \leq t \leq t_{n+1} = t_n+T_s$ with $n = 0, \dots, N_{c}-1$, with $T_s = T/N_{c}$ being the interval between inputs (sampling time). We define the input rate of movement as $\Delta u_n = u_n-u_{n-1}$ and we also impose input bound constraints. This constitutes the OCP
\begin{subequations}
\label{eq:OCP}
\begin{alignat}{2}
       &\min_{x, u} \quad &&\phi = \phi_{z} +  \phi_{\Delta u},\\
       &s.t.
       &&x(t_0) = x_0,\\
       & &&\dot{x}(t) = f(t, x(t), u(t), d(t)),\; t \in [t_0,\, t_f],\\
       & &&z(t) = h(t, x(t), u(t), d(t)),\; t \in [t_0, \,t_f],\label{subeq:output_equation}\\
       & &&u(t) = u_{n},\;  t_n \leq t \leq t_{n+1}, \; n \in \mathcal{N},\\
       & &&u_{\min} \leq u(t) \leq u_{\max}, t\in [t_0,\, t_f],
\end{alignat}
\end{subequations}
with $\mathcal{N} = 1, \dots, N_{c}-1$, and the objectives
\begin{subequations}
\label{eq:objectives}
\begin{alignat}{2}
       \phi_{z} &= \frac{1}{2}\int_{t_0}^{t_f}\left \|  z(t)-\Bar{z}(t)\right \|_{Q_{z}}^2dt,\\
       \phi_{\Delta u} &= \frac{1}{2}\sum_{j=0}^{N_{c}-1}\left \|  \Delta u_{j}\right \|_{\bar{Q}_{\Delta u}}^2,
\end{alignat}
\end{subequations}
where $z(t)$ is the output, $\phi_{z}$ penalizes the difference between the output $z(t)$ and the setpoint $\Bar{z}(t)$ and $\phi_{\Delta u}$ penalizes the input rate of movement. $Q_{z}$ and $\bar{Q}_{\Delta u} = Q_{\Delta u}/T_s$ represent the weight matrices for $\phi_{z}$ and $\phi_{\Delta u}$, respectively.
We transcribe the OCP \eqref{eq:OCP}-\eqref{eq:objectives} into a nonlinear programming (NLP) problem using a multiple shooting approach. We solve an IVP between each input interval using the ESDIRK methods with $N$ steps. To solve the NLP, we apply an SQP with a line-search algorithm and BFGS to update the Lagrange Hessian developed in \cite{Keysfeld2023}.

\section{NUMERICAL EXPERIMENTS}
\label{chap:Numerical_experiments}

We compare the computational performance of the iterated and direct approaches in ESDIRK-based OCP. We do this by repeatedly solving OCPs with a varying number of integration steps between control intervals for the model of the QTS presented in \cite{Andersen2023}. We compare these two approaches to a base case that applies an ESDIRK method with iteration matrix refactorizations at every Newton-type iteration. This base case applies a direct approach which due to the frequent refactorizations, achieves accurate sensitivity information. Table \ref{tab:QTS_parameters_and_initial_conditions} presents the model parameters for the QTS. We use the initial conditions $x(t_0)= [7602.7, \,11404.0, \,1000.0, \,1000.0]'$ g and apply the constant vector of disturbances $d(t) = [0,\, 0,\, 100,\, 100]'$ cm$^3$/s for all experiments. The outputs of the QTS are the water levels in the two bottom tanks, i.e., $z(t) = \begin{bmatrix}
\frac{1}{\rho A_1}x_1(t), & \frac{1}{\rho A_2}x_2(t)
\end{bmatrix}'$.

\begin{figure}[tb]
    \centering
    \includegraphics[width=0.48\textwidth]{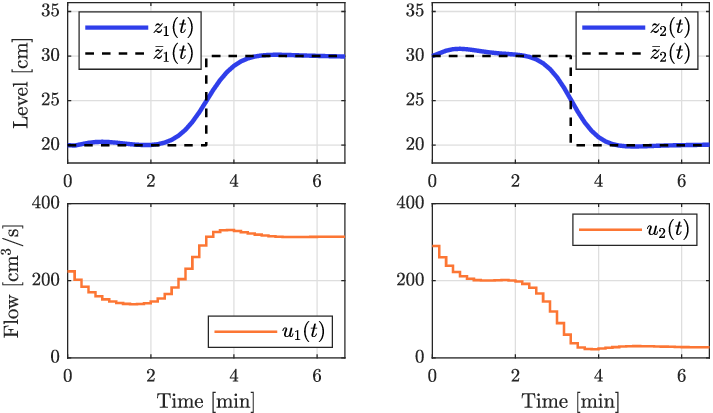}
    \caption{A solution of the OCP applying ESDIRK12 with $N=10$ and iterated IND sensitivities.}
    \label{fig:simulation_example}
\end{figure}

For the OCP, we choose the sampling time $T_s = 10$ s. We use the weight matrices $Q_z = \text{diag}([10,\, 10])$ and  $Q_{\Delta u} = \text{diag}([0.1,\, 0.1])$, and the number of steps in the horizon $N_{c} = 40$. The bounds are between 0 cm$^3$/s and 500 cm$^3$/s for both inputs. For the experiments, we construct the time-varying setpoints $\Bar{z}(t) = \begin{bmatrix}
20, & 30
\end{bmatrix}'$ cm for $0 \leq t < T/2$ and $\Bar{z}(t) = \begin{bmatrix}
30, & 20
\end{bmatrix}'$ cm for $T/2 \leq t \leq T$. We initialize all states and inputs for the SQP as a vector with all entries set to 300. For the SQP algorithm, we choose the SQP tolerances as $10^{-3}$, the QP tolerance $10^{-8}$, the SQP step
length tolerance $10^{-8}$, and abs and rel in \eqref{eq:stop_crit_newton}   both as $10^{-8}$. Fig. \ref{fig:simulation_example} presents one solution of the OCP using the ESDIRK12 with 10 integrator steps between control intervals and iterated IND sensitivity computations.

\begin{table}[tb]
\centering
\caption{Parameters for QTS for $i=1,\dots, 4$.}
\label{tab:QTS_parameters_and_initial_conditions}
\begin{tabular}{ccccc}
      $a_i$ [cm$^2$] & $A_i$ [cm$^2$]   & $\gamma_{\text{valves}}$ [-] & $\rho$ [g/cm$^3$] & $g$ [cm/s$^2$] \\ \hline
$1.2272$ & $380.1327$ & $[0.6; \,0.7]$  & $1.0$ & $981$
\end{tabular}
\end{table}

\subsection{Varying integration step-size between control intervals}
\begin{figure}[tb]
    \centering
    \includegraphics[width=0.48\textwidth]{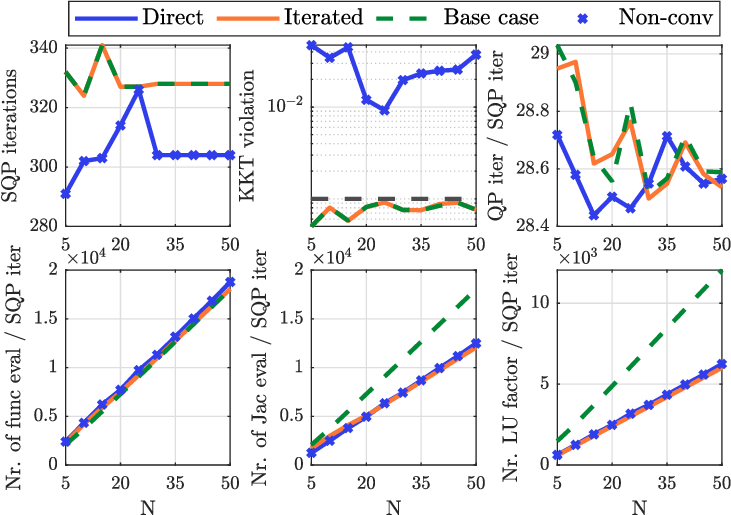}
    \caption{OCP statistics using ESDIRK12.}
    \label{fig:ESDIRK12_sweep}
\end{figure}

\begin{figure}[tb]
    \centering
    \includegraphics[width=0.48\textwidth]{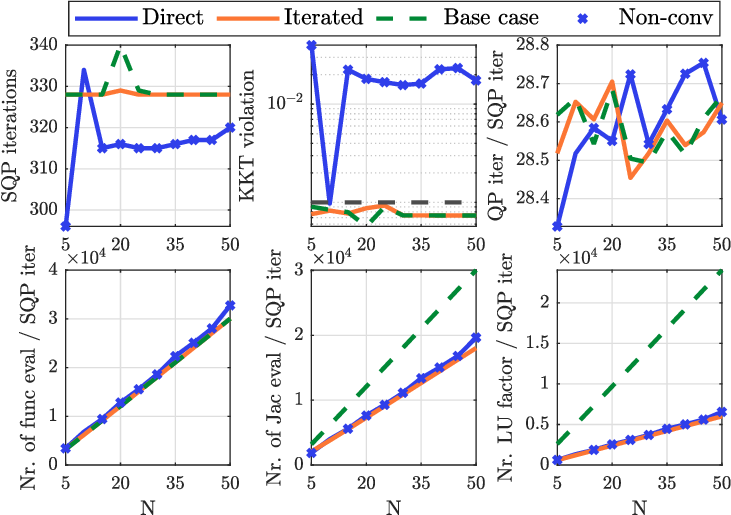}
    \caption{OCP statistics using ESDIRK23.}
    \label{fig:ESDIRK23_sweep}
\end{figure}

\begin{figure}[tb]
    \centering
    \includegraphics[width=0.48\textwidth]{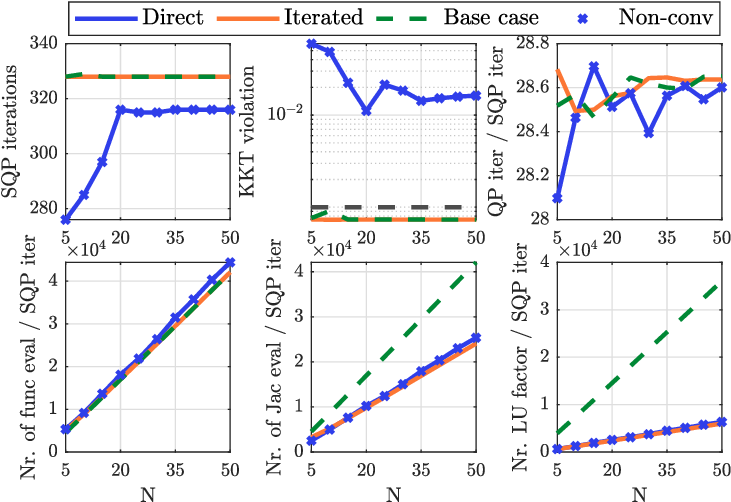}
    \caption{OCP statistics using ESDIRK34.}
    \label{fig:ESDIRK34_sweep}
\end{figure}

We solve the OCPs repeatedly while increasing the number of integration steps between control intervals. We apply the sequence $N = [5,\, 10,\, 15,\, 20,\, 25,\, 30,\, 35,\, 40,\, 45,\, 50]$ of integration steps between controls. We also do this with the base case. To measure the performance of the three different cases, we store the number of SQP iterations, the KKT violations, and the number of QPs iterations for the optimizer. We also store the total number of function and Jacobian evaluations, and LU factorizations of the iteration matrix.
Fig. \ref{fig:ESDIRK12_sweep}, \ref{fig:ESDIRK23_sweep}, and \ref{fig:ESDIRK34_sweep} present the results using ESDIRK12, ESDIRK23, and ESDIRK34, respectively. The dotted black lines are the SQP tolerance representing the allowed KKT violations.
The crosses in these figures represent experiments that do not converge. The non-convergence is due to the step length, computed using the line-search algorithm,  exceeding the step tolerance.

\subsection{Low SQP tolerance and short control intervals}

\begin{figure}[tb]
    \centering
    \includegraphics[width=0.48\textwidth]{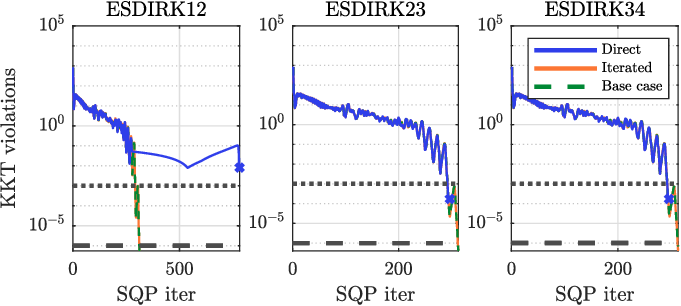}
    \caption{Comparison of ESDIRK-based OCP algorithms with direct and iterated sensitivity computations. All methods apply 10 integration steps between control intervals..}
    \label{fig:low_tolerance_experiment}
\end{figure}

We also demonstrate the performance when we lower the tolerances for the SQP to $10^{-6}$, the QP to $10^{-10}$, and abs and rel to $10^{-10}$. We also choose a smaller control interval at $T_s = 2.0$ s. We keep the same step tolerance.  Fig. \ref{fig:low_tolerance_experiment} shows the results. Crosses represent experiments that do not converge to the SQP tolerance of $10^{-6}$.

\subsection{Comments on the numerical experiments}

Fig. \ref{fig:ESDIRK12_sweep}-\ref{fig:ESDIRK34_sweep} demonstrate that ESDIRK methods using the iterated sensitivity approach converge to the desired SQP tolerance of $10^{-3}$ in all experiments. The direct approach only achieves convergence for ESDIRK23 when using 10 integration steps between control intervals. Additionally, the iterated approach performs similarly to the base case while requiring fewer total Jacobian updates and LU factorizations. The base case requires significantly more LU factorizations compared to the iterated approach as it refactorizes the iteration matrix at every Newton iteration.  Both the base case and the iterated approach update the Jacobians once per Newton iteration, but the base case requires one additional Jacobian update before every Newton scheme. It requires this, as it computes the Jacobian for the explicit state and also for constructing the iteration matrix based on the SVPs before every Newton scheme.
In contrast, the iterated approach uses the same Jacobians for both the explicit step and the iteration matrix. Finally, Fig. \ref{fig:low_tolerance_experiment} demonstrates that the direct method can achieve convergence for ESDIRK23 and ESDIRK34 as long as we maintain an SQP tolerance of $10^{-3}$ and a short control interval. However, it still cannot reach the low tolerance requirement at $10^{-6}$.

\section{CONCLUSIONS}
\label{chap:Conclusions}
In this paper, we compare the iterated and direct approaches to sensitivity computation for ESDIRK-based OCPs using multiple shooting discretization. The iterated approach strictly applies the principle of internal numerical differentiation, and the direct method approximates sensitivities without employing the Newton-type schemes from the ESDIRK methods. We evaluate both methods in terms of some computational performance metrics by repeatedly solving ESDIRK-based OCPs for the QTS with a varying number of integration steps between control intervals. The computational performance metrics are the number of SQP and QPs iterations, KKT violations, the total number of function evaluations, Jacobian updates, and iteration matrix factorizations. We benchmark these approaches to an ESDIRK method that refactors the iteration matrix in every Newton-type iteration and employs a direct sensitivity computation approach. We refer to this as the base case and the study shows that the iterated method performs similarly to the base case but requires fewer Jacobian updates and matrix factorizations. Furthermore, the iterated approach outperforms the direct method for longer sampling times.

\bibliographystyle{IEEEtran}

\bibliography{ref/references}

\end{document}